\input harvmac
\noblackbox
\font\ticp=cmcsc10
 
\def\Title#1#2{\rightline{#1}\ifx\answ\bigans\nopagenumbers\pageno0\vskip1in
\else\pageno1\vskip.8in\fi \centerline{\titlefont #2}\vskip .5in}

\font\ticp=cmcsc10
\font\ttsmall=cmtt10 at 8pt

%
%
\def\[{\left [}
\def\]{\right ]}
\def\({\left (}
\def\){\right )}
\def\sh{ \sinh}
\def\ch{ \cosh}
\def\th{ \tanh}
\def\a {\alpha}
\def\apm{\alpha'}
\def\t{\theta}


\lref\su {
L. Susskind, hep-th/9309145.}
\lref\hrs{ E. Halyo, A. Rajaraman and L. Susskind, hep-th/9605112.}
\lref\hkrs{
E. Halyo, B. Kol, A. Rajaraman and L. Susskind, hep-th/9609075.}
\lref\pol{ J. Polchinski, Phys. Rev. Lett. {\bf 75} (1995) 4724,
hep-th/9510017. }
\lref\hlm {G. Horowitz, D. Lowe, and J. Maldacena, Phys. Rev. Lett.
{\bf 77} (1996) 430, hep-th/9603195.}
\lref\ghjp{ G. T. Horowitz and J. Polchinski, Phys.
Rev. {\bf D55} (1997) 6189, hep-th/9612146.}
\lref\ghjpII{ G. T. Horowitz and J. Polchinski, 
 hep-th/9707170.}
\lref\hs{H. Sheinblatt, hep-th/9705054.}
\lref\rb{
R. Emparan, hep-th/9704204;
S. Das, hep-th/9705165;
S. Mathur, hep-th/9706151;
 S. Das, S. Mathur, S. Rama and P. Ramadevi, hep-th/9711003;
G. T. Horowitz and J. Polchinski,
 hep-th/9707170. }
\lref\ts{A. Tseytlin, Nucl. Phys. {\bf B475} (1996) 149,
hep-th/9604035;
I. Klebanov and A. Tseytlin, Nucl. Phys. {\bf B475}
(1996) 179,
hep-th/9604166;
M. Cvetic and A. Tseytlin, Nucl. Phys.{\bf B487}(1996) 181,
hep-th/9606033 }

\lref\bmm{J. Breckenridge, G. Michaud and R. Myers, Phys. Rev.
{\bf D55} (1997) 6438, hep-th/9611174.}
\lref\jmas{J. Maldacena and A. Strominger, Phys. Rev. Lett.
{\bf 77} (1996) 428, hep-th/9603060.}
\lref\ghas{
C. G. Callan, Jr. and J. M. Maldacena, Nucl. Phys. {\bf B472}
(1996) 591, hep-th/9602043;
G. T. Horowitz and A. Strominger, Phys. Rev. Lett. {\bf 77}
(1996) 2368, hep-th/9602051}

\lref\hms{G. T. Horowitz, J. M. Maldacena, and
A. Strominger, Phys. Lett. {\bf B383} (1996) 151, hep-th/9603109.}
\lref\vbfl{ V. Balasubramanian and F. Larsen, Nulc. Phys. {\bf B478}(1996)
199,
hep-th/9604189. }

\lref\nojz{N. Ohta and J. Zhou, hep-th/9706153}
\lref\gl{G. Lifschytz, Nucl. Phys. {\bf B499} (1997) 283,
hep-th/9610125.}
\lref\more{See e.g. G. Gibbons and K. Maeda, Nucl. Phys. {\bf B298} (1988) 741;
P. Bizon, Phys. Rev. Lett. {\bf 64} (1990) 2844; K. Y. Lee, V. P. Nair and
E. Weinberg, Phys. Rev. Lett. {\bf 68} (1992) 1100.}
\lref\pois{E. Poisson and W. Israel, Phys. Rev. {\bf D41} (1990) 1796.}
\lref\ori{A. Ori,  Phys. Rev. Lett. {\bf 67} (1991) 789; {\bf 68} (1992) 2117.}
\lref\ghdm{G. Horowitz and D. Marolf, Phys. Rev. {\bf D55} (1997) 835,
hep-th/9605224.}
\lref\ghd{G. Horowitz and D. Marolf, Phys. Rev. {\bf D55} (1997) 846,
hep-th/9606113.}
\lref\bek{J. Bekenstein, gr-qc/9605059.}
\lref\lawi{F. Larsen and F. Wilczek, Phys. Lett. {\bf B375} (1996) 37, 
hep-th/9511064; Nucl. Phys. {\bf B475} (1996) 627, hep-th/9604134;
hep-th/9609084.}
\lref\cvts{M. Cvetic and A. Tseytlin, Phys. Rev. {\bf D53} (1996) 5619,
hep-th/9512031; A. Tseytlin, Mod. Phys. Lett. {\bf A11} (1996) 689,
hep-th/9601177;  Nucl. Phys. {\bf B477} (1996) 431, hep-th/9605091.}
\lref\kmr{N. Kaloper, R. Myers and H. Roussel, hep-th/9612248.}
\lref\ascv{A. Strominger and C. Vafa, Phys. Lett. {\bf B379} (1996) 99,
hep-th/9601029.}
\lref\homa{G. Horowitz and D. Marolf, hep-th/9610171.}
\lref\jp{J. Polchinski, Phys. Rev. Lett. {\bf 75} (1995) 4724,
hep-th/9510017.}
\lref\mtw{C. Misner, K. Thorne, and J. Wheeler, {\it Gravitation}, Sec. 32.6
(W. H. Freeman, New York, 1973).}
\lref\gar{D. Garfinkle and T. Vachaspati, Phys. Rev. {\bf D42} (1990) 1960;
D. Garfinkle, Phys. Rev. {\bf D46} (1992) 4286.}
\lref\host{G. Horowitz and A. Steif, Phys. Rev. Lett. {\bf 64} (1990) 260.}
%
\baselineskip 16pt
\Title{\vbox{\baselineskip12pt
\line{\hfil   UCSBTH-98-01}
\line{\hfil \tt hep-th/9801074} }}
{\vbox{
{\centerline{Black Holes with Multiple  Charges and}
\centerline{ the Correspondence Principle 
}}
}}
\centerline{\ticp Haisong Yang\foot{\ttsmall E-mail:
yangh@cosmic1.physics.ucsb.edu} }
\bigskip
\vskip.1in
\centerline{\it Department of Physics, University of California,
Santa Barbara, CA 93106, USA}
\bigskip
\centerline{\bf Abstract}
\bigskip
We consider the entropy  of near extremal
 black holes with multiple charges
in the context of 
 the recently proposed correspondence principle of Horowitz and
Polchinski, including black holes with two, three and four Ramond-Ramond
charges.  
 We find that at the 
matching point the black hole entropy can  be accounted for by massless
open strings ending on the D-branes for  all cases except a black hole
with four Ramond-Ramond charges, in which case a possible resolution in
terms of brane-antibrane excitations is considered.
\Date{January, 1998}

\newsec{Introduction}

A few years ago, Susskind proposed \su\     that there should be
a one-to-one correspondence  between quantum states of a Schwarzschild
black
hole and fundamental string states. This idea was carried out further
in \hrs \hkrs. However, it was believed that there was a discrepancy
between these two objects: the black hole entropy scales as $M^2$ while
the entropy of a fundamental string scales as $M$. 
Recently, Horowitz and Polchinski formulated a  correspondence principle
\ghjp\ for 
black holes 
and strings 
   which resolves the discrepancy in a certain way.
 It states that quantum
states of black holes can be matched onto  states of
strings and D-branes \jp\  at a special point of the string
coupling constant $g$ and during the transition the mass 
changes by at most a factor of order  unity.

More precisely, the correspondence principle works in the following way.
 Start  
with the black hole picture. Decrease $g$ until 
the curvature (in the string metric)
at the event horizon  
becomes of order  the string scale. 
At this point (called the matching point),
 the $\apm$ corrections in  the low energy effective  
action  become   
important and the black hole solution can not be trusted any more. So,
 the black hole is gone and the physical picture changes, what 
 takes the place of the black hole is  
some  description in terms of strings and D-branes,
since these are the dynamic degrees of freedom that 
we have. On the black hole side there is only  a thermodynamic 
description, while on   
 the string side there is  
 also a statistical description, which  enables one to 
calculate the entropy microscopically.       
What was found is that the thermodynamic
quantities, such as entropy, temperature, etc., of these two
pictures agree with each other.
 Of course, 
we do not have a  precise description of the transition regime,
so the matching is only up to factors of order 
unity. 

In \ghjp, three different cases were considered:
Schwarzschild black hole, near extremal black holes with fundamental
string winding and momentum charge, near extremal black p-brane with 
a single  RR(Ramond-Ramond)  charge.  Agreement
was found for all three cases. Notice that for the last two cases, if 
the black holes are not near extremal, i.e., if the charges that they
carry are small, 
then they are qualitatively the same as a Schwarzschild black hole, 
since the correspondence principle is not sensitive to factors of order
unity. 

At the matching point, 
the Schwarzschild black
hole
 is described by  a long self-interacting fundamental string 
that has  size of order  $\apm^{1/2}$. 
  The second black hole is described by
an excited fundamental
string that is wrapped around a  $S^{1}$ and also carries
some momentum in the same direction. The black p-branes are described
by
D p-branes  wrapped around the internal
torus with some excitations  described by massless 
open strings ending on the D-branes. It was thought in \ghjp\ that 
 there is a large gravitational
 field dressing for all $p$-branes at the matching point.   We
 will show  that while this is true for  $p=4, 5, 6$,
 there is no large dressing for
$p=0, 1, 2, 3$. 

It is natural to ask whether the correspondence principle
 can be applied 
to other black holes. An important class of black holes are 
the ones that carry multiple RR charges in type IIA or
type IIB string theory. If the correspondence principle can be 
applied to them,  then at  the matching point, they should all 
be described by certain D-brane configurations with some excitations.
   
D-branes  have even
spatial dimensions in type IIA and odd spatial dimensions in type IIB. Upon
compactification on a torus, T-duality 
 interchanges type IIA and type IIB, even dimensional
D-branes  and odd dimensional D-branes. If we have two kinds of 
D-branes, the number of ND (Neumann-Dirichlet) type of boundary 
conditions between them, however,
 is invariant under T-duality. So  black holes that 
carry two RR charges  can be put  into the following 
 three categories according to the 
the number of ND conditions: 

\ \ (A) 2 ND conditions 

\ \ (B) 4 ND conditions 

\ \ {(C) 6 ND conditions.

 A black hole with D 0- and D 6-brane charges 
was examined  in 
\hs\ in the context of the correspondence principle  and agreement was found. In this paper, we consider all  black holes
in    category (A) and (B). We also consider  black holes with
three  or four RR charges.  

The plan of this paper is the following. In section II we examine 
the issue of gravitational field dressing of the black p-branes with a
single RR charge. In section III we first consider a five dimensional black
hole with D 1- and D 5-brane charges and  a six dimensional black hole
with D 0- and D 4-brane charges, both of which belongs to category (B).
We also include the discussion of 
 a four dimensional black hole that carries D
2-brane, D 6-brane and NS(Neveu-Schwarz)  5-brane charges because it shares some common
feature with the above two black holes. In section IV, we consider
a four dimensional black hole that can carry up to four RR charges. It is a
black hole in  category (B)  
if  two of the charges are set to be zero.
We also consider
the case when three or four charges are large. In section V, we consider
an eight dimensional black hole that carries  
 D 0- and D 2-brane charges, which  belongs to   
 category (A). We  also discuss a generalization of this 
black hole.
We conclude in section VI. 
 
Other aspects of the     
correspondence principle
were examined in \rb.

\newsec{Black Holes with a Single RR Charge}

In this section, we consider the black $p$-branes with a single RR charge. 
   In \ghjp\ it was noticed that if one starts with D-branes in flat
spacetime at zero string coupling and increases the coupling constant,
then gravitational effects become important before the black hole is
formed. In particular the locally measured 
size of the internal torus and temperature can differ from their
asymptotic values by a large factor.  We 
  show  that while  this large gravitational
 field dressing  is indeed present for  $p=4, 5, 6$, 
 there is no large gravitational dressing for
$p=0, 1, 2, 3$. 
 The conclusion that the  entropy can be accounted for by the
massless open strings however is still correct, 
  as we will see.

The string metric of the black p-branes is given by 
\eqn\blackpbrane{
ds^2 = H^{-1/2}\[-\(1-{r_0^n \over r^n}\)
dt^2 + dy^a dy_a\] 
+ H^{1/2} \[\(1-{r_0^n \over r^n}\)^
{-1} dr^2 + r^2 d\Omega_{n+1}^2\] } 
where 
\eqn\pbranehmfn{
H = 1 + {r_0^n \sh^2\a \over r^n}}
The $y_a$ are $p=7-n$ spatial coordinates along the brane which we
assume  are compactified on a torus. The dilaton is $e^{2\phi}= H^{(n-4)
/ 2}$. The horizon is at $r=r_0$. 

Let us define    $r^* =  r_0 \(\ch\a\)^{2/n}$,  the point where $H$
becomes of order 1. The crucial thing to notice is that the spatial
geometries outside the
horizon are different  for different $p$-branes.
The angular part of metric is   $\(1 + {r_0^n \sh^2\a \over r^n
}\)^{1/2} r^2 d\Omega^2_{n+1}$. 
If $n < 4\  (p=4, 5, 6)$, the radius of the $S_{n+1}$ increases with some power in
$r$ outside the horizon. If $n = 4\  (p=3)$, the radius of the $S_5$ does not change much for $r_0
< r < r^*$ and  increases linearly with $r$ for $r> r^*$.
 In other words,  there is a deep throat  from $r_0$ to $r^*$.
If $n > 4\ (p=0, 1, 2)$, the radius of the $S_{n+1}$ first decreases, it
reaches the  
minimum  at $r \sim r^*$ and then increase linearly with $r$.
We will refer
to this kind of  geometry as the ``bottle neck''.

By the  correspondence principle the matching between  black holes and
D-branes and strings happens when the curvature at the horizon becomes
of order the string scale. For these black p-branes, the largest
contribution to the curvature 
comes from the angular part of the metric. This means   the
matching happens when the radius of the $S_{n+1}$ is of order
$\apm^{1/2}$.
 We immediately see the difference between different $p$-branes. 
 For
 $p = 4, 5, 6 $, the stringy
behavior only smears out to $r \sim r_{0}$ ( the zero in
$\(1-{r_0^n \over r^n}\)$ is  smeared out). So a large gravitational
field   dressing is left and 
the local temperature is of order the string scale due to the redshift.
  For $p = 3$, 
 the stringy behavior not only smears out the horizon, but
also the whole throat and the metric \blackpbrane\ can only be trusted for $r
> r^*$. As a result, there is no large 
dressing for the 3-brane and  the spacetime is basically flat.  Similarly, for $p = 0, 1, 2$, the stringy behavior smears out the metric 
all the way to $r \sim r_0 (\ch\a)^{3 \over n+2}$, where the curvature
becomes of order the string scale again. We can imagine the process of
approaching the matching point from the black hole side for the last case. The 
fundamental string length  
 grows relative to the size of the black hole as we decrease
the coupling constant. The stringy
behavior should first smears out the region at the minimum radius $r \sim r^*$, the geometry there gets corrected but the black hole still exists.
As the coupling constant is further decreased, the smeared out region
will grow and eventually, at the matching point, the horizon is smeared
out. 
 Since there is no large dressing for $p \le 3$, the local
temperature is the same as asymptotic temperature and is small in string
units.

The above observation  nevertheless does not jeopardize the conclusion that
 the entropy of all $p$-branes  
 can be accounted for by  the  massless open strings on the D-branes.
A priori, we should do the calculation locally at the D-branes. However, 
 it was  pointed out in \ghjp\ 
that in fact one  
 can do the  calculation either locally or at $r =
\infty$, which  follows from the observation that    
the entropy and energy expressions of the ideal gas remain the same under
the rescaling due the large dressing. We have just seen that there is no
large dressing  for black
p-branes with $p \le 3$, 
 so we should use
the asymptotic quantities to do the calculation for them anyway. The
conclusion   therefore is unaffected. We should remark that this is the
special property of  black holes with a single RR charge. For
black holes with more than one charge, in general we can only do the
calculation locally.

 \newsec{Three Black Holes with Two or Three Charges}
In this section, we consider three different black holes: a five dimensional
black hole with D 1- and D 5-brane charges, a six dimensional black hole
with D 0- and D 4-brane charges and a four dimensional black hole with D
2-brane, D 6-brane and NS 5-brane charges. All charges are assumed to be
large.
Although the precise counting of entropy 
 has already been done for the first and
third black hole (but  not for the second one), 
we want to apply the correspondence principle to all of them 
as a test for the correspondence principle. We will see that it can be
applied to  all three  black holes successfully. In  particular, we
resolve a puzzle discussed in \ghjp. 

We consider these black holes together in this section because 
they share some common features.  For all three black holes,
the spatial geometry outside the 
horizon is either a deep throat or a ``bottle neck", which are  
smeared out by the stringy behavior at the matching point.  This
reduces the redshift effect and causes  the local temperatures  
 to be  small in string scale. These features are   similar to
the black p-branes with $p \le 3$ discussed in the last section. On the
other hand, these black holes carry more than one charge, so there could
still be large gravitational field dressing, which depends on the relative
magnitudes of the boost parameters. This will become clear in later
discussions. 

In the following, we first consider the five dimensional black
hole with D 1- and D 5-brane charges in detail in 3.1. We then briefly discuss 
the other two black holes in 3.2 and 3.3, in particular the new features of each one.

\subsec{Five Dimensional Black Hole with D 1- and D 5-brane Charges}
The string metric of the five dimensional black hole with D 1- and D
5-brane charges is given by
\eqn\didvmetric{\eqalign{
ds^2 = &(H_1 H_5 )^{1/2} \Bigl[ -(H_1 H_5 )^{-1} f dt^2
+ (H_1 H_5)^{-1}  dy_1^2 +  H_5^{-1} (dy_2^2 +
 dy_3^2  +  dy_4^2 +
 dy_5^2)  \cr
&+ f^{-1}dr^2 +r^2 d\Omega_3^2 \Bigr] }}
where
\eqn\didvhmfn{\eqalign{
H_i  &= 1+{r^2_0 \sinh^2\alpha_i \over r^2} \ \ (i=1, 5) \cr
f &= 1-{r^2_0 \over r^2} }}
The $y_a$ are coordinates
on the internal $T^5$ which are
identified with length $L_a$.  The dilaton is
$e^{2 \phi} = H_1 H_5^{-1}$

This black hole is a solution to the type
IIB string theory on $T^5$ and can be constructed by
the `harmonic function' rule described in \ts.
 The  horizon is at $r=r_0$.
The extremal limit of the black hole can be obtained
by letting $r_0 \to 0$  while fixing $r^2_0 \sh^2\a_i$ 
 and has zero horizon area. 
The extremal limit   preserves 1/4 of the 
supersymmetries and can be  
thought of as  a superposition of 
extremal black 5-branes wrapped around the $T^5$ 
 and black 1-branes wrapped along the  $y_1$ direction and  smeared out 
in the rest four internal directions.
The black hole is near extremal when at least
one  RR charge is large (i.e., one $\ch\a_i$ is much larger than 1).
 Near extremal black holes
can be thought of as excited states of the extremal limit. If only one
charge is large, the black hole is essentially the same as the black
holes with a single charge, since the correspondence principle would not
be able to distinguish them.   
  We will consider the case when both charges are large.

The energy, excess energy, entropy, Hawking temperature and two RR charges
of the
black hole are
 \eqn\didvpara{\eqalign{
 E &\sim {L_1 L_2 L_3 L_4 L_5  \over g^2 \apm^4}\  r_0^2
 (\ch 2\alpha_1 + \ch 2\alpha_5  + 1) \cr
 \Delta E_{bh} &\sim {L_1 L_2 L_3 L_4 L_5  \over g^2 \apm^4}\  r^2_0
   \cr
 S_{bh} &\sim {L_1 L_2 L_3 L_4 L_5  \over g^2 \apm^{4}} \ r_0^3  \ch
\alpha_1 \ch\alpha_5  \cr
 T &\sim {1 \over  r_0 \ch\alpha_1 \ch\alpha_5 }
\cr
 Q_1 &\sim { L_2 L_3 L_4 L_5 \over g \apm^3} \ r_0^2 \sh 2\alpha_1 \cr
 Q_5 &\sim { 1 \over g \apm} \ r_0^2 \sh 2\alpha_5 \cr
    }}
The excess energy $\Delta E_{bh}$ is the energy above the
BPS limit. We have dropped the overall numerical coefficients in
 these expressions
since the correspondence principle  is not
sensitive to them. However, the coefficients in
front of $E$ and $Q_i$ are the same, so that
 $\Delta E_{bh}$ takes the above form. 
 For
 near BPS black holes, the
correspondence principle states that the excess energy changes by at
most a factor of unity during the transition.
 This will be used when we calculate the entropy
on the string side.

The matching between the black hole states and states
of D-branes and strings occurs  when the curvature
 at the horizon is of order $1/ \apm$. The largest contribution 
comes from the angular part of the metric and is of order $(r_0^2
\ch\a_1 \ch\a_5)^{-1}$. 
 So the
matching point is
\eqn\didvmpt{
r_0 \sim {  \apm^{1/2} \over (\ch\a_1 \ch\a_5)^{1/2}}}

 Let us define $r^* = r_0 \ch\a_{min}$, where $\ch\a_{min}$ is the smaller
one of $\ch\a_1$ and $\ch\a_5$. The spatial geometry outside the horizon
 has a deep throat from $r_0$ to $r^*$, which is smeared out by
stingy behavior at the matching point. If $\ch\a_1 \sim \ch\a_5$, 
we have $H_1(r=r^*) \sim H_5(r= r^*) \sim 1$ and there is no large gravitational
field dressing;
if there is a hierarchy between $\ch\a_1$ and $\ch\a_5$, then there is still a
large dressing. In either case, we can write 
$H_i(r=r^*) \sim {\ch^2\a_i \over 
\ch^2\a_{min}}, (i = 1, 5)$. In the following, we will treat both cases
together, all formulas will be be valid for both. 
It is instructive to write down the metric at the D-branes
(i.e., $r=r^*$):
\eqn\didvlocalmetric{\eqalign{
ds^2 \sim\  &\({\ch\a_1 \ch\a_5 \over \ch^2\a_{min}}\)
 \Biggr[- {dt^2 \over \({\ch\a_1  \ch\a_5 \over
\ch^2\a_{min}
}\)^2} 
+{dy_1^2 \over \({\ch\a_1 \ch\a_5 \over \ch^2\a_{min}}\)^2} \cr
&+  {dy_2^2+ dy_3^2 + dy_4^2 + dy_5^2 \over 
\({\ch\a_5  \over \ch\a_{min}}\)^2} + 
\({\ch^2\a_{min}\over \ch\a_1  \ch\a_5}\) \apm d\Omega_3^2  
 \Biggr] }}
where we have substituted \didvmpt\ into the metric.
Let us use $T',\Delta E_{bh}', L_a'$ to
denote the local temperature, excess energy
and  spatial sizes of the
$T^5$. Following \didvpara\ \didvmpt\  \didvlocalmetric, they are given
by 
\eqn\didvlocalpara{\eqalign{
T' &\sim { 1 \over \apm^{1/2}\ch\a_{min}} \cr
\Delta E_{bh}' &\sim {L_1 L_2 L_3 L_4 L_5    \over g^2 \apm^{3}  \(\ch\a_1
\ch\a_5\)^{1/2} \ch\a_{min}} \cr 
L_1' &\sim L_1 { \ch\a_{min}  \over \(\ch\a_1 \ch\a_5 \)^{1/2}}\cr 
L_i' &\sim L_i \({\ch\a_1 \over \ch\a_5}\)^{1/2} (i=2, 3, 4, 5)
}}
These expressions are correct independent of whether there is large
 dressing or not. At the matching point, the entropy and charges are
\eqn\didvmpsc{\eqalign{
S_{bh} &\sim {L_1 L_2 L_3 L_4 L_5  \over g^2 \apm^{5/2}
\(\ch\a_1 \ch\a_5\)^{1/2}} \cr
Q_1 &\sim { L_2 L_3 L_4 L_5 \ch\a_1\over g \apm^2 \ch\a_5} \cr
  Q_5 &\sim { \ch\a_5 \over g \ch\a_1} 
    }}                                                              

Now let us consider the statistical description of the system. It should
be described by  $Q_1$ D 1-branes and $Q_5$ D
5-branes with some excitations. 
We will    assume   the excitations are described by massless open
strings
on the D-branes
 and see that it   gives consistent result.
As we discussed at the end of section 2, the calculation has to be done
locally at the D-branes.

There are three kinds of open strings : 1-1, 5-5 and 1-5  strings.
For fixed amount of excess energy $\Delta E'_{bh}$, one can ask
the following question: to  maximize the  entropy,
what is the correct  distribution
of  $\Delta E'_{bh}$ among the three kinds of open strings.
The answer is that  the three kinds of open
strings should have the same temperature.
It turns out that  if we put $\Delta
E'_{bh}$ into the open strings, we get a local temperature that is   
   the same  as  
the $T'$ in \didvlocalpara.  This is part of the matching property  that we  would like
to have.  
To facilitate the
calculation, we will  just assume the local temperature to be
$T'$ and
 calculate the entropy and energy carried by these open strings. 
We will   see that
they agree with $S_{bh}$ and $\Delta E_{bh}'$.

Let us use  $S_{ij}$ to denote
the entropy contribution from the $i$-$j$ strings. We consider the 5-5
strings first. The 5-5 strings are a five 
dimensional  gas at local temperature $T'$. 
     The number of species
of open strings  is
$Q_5^2$. So the entropy is 
\eqn\didvsvv{
S_{55}  \sim \  Q_5^2
  L_1' L_2' L_3' L_4' L_5' T'^5
\sim {S_{bh} \over \ch^4\a_{min}}}
The second step follows from direct substitutions of \didvlocalpara\ and
\didvmpsc.
 The above picture is not
valid if $T'L_i'<1$ for any
$L_i'$,   since the open strings do not see
those dimensions and behave like a lower dimensional gas. In such case,
we should use the wrapped picture to calculate the entropy, i.e., let
the D-branes be wrapped around those small circles many times(See \ghjp\ for
detailed discussions). This will 
 increase the effective volume of the gas and reduce the number of
species of strings. The net result is that equation \didvsvv\ is still valid
and gives the correct entropy. 
 
Next, we  consider 1-1 strings. We can avoid the calculation
if we notice the T-dual relationship 
between the 1-1 and 5-5 strings. If we T-dualize the $y_2, y_3, y_4, y_5$
directions, the 5-branes become 1-branes and 5-5 strings become 1-1
strings and vice versa. But the $S_{55}$ expression ${S_{bh} \over
\ch^4\a_{min}}$
is invariant under this operation, so we must have $S_{11} \sim
S_{55}$. 

Although not necessary, it is still
  instructive to  calculate $S_{11}$ directly in the original
picture. We need to first determine the number of species of 1-1 strings
that are excited. The $Q_1$  D 1-branes
 are evenly distributed in the $y_2, y_3, y_4, y_5$ directions.
Let us focus on the $y_2$ direction for now.
  Because
the local temperature is of order $T'$,
the open strings that are excited should
have  length of  order  $\apm T'$ or
smaller. An open string that starts from a
particular  1-brane can end on any  1-brane within
$\apm T'$ distance, so the number of 1-branes that it can end on is
$Q_1{\apm T'
\over L_2'} $. The total number of 1-1 strings that are excited is
therefore $Q_1^2{\apm T'
\over L_2'} $.  We can also try to understand the factor ${\apm T'
\over L_2'} $ in the T-dual picture. If we T-dualize the $y_2$
direction, the 1-branes becomes 2-branes.
 We have  a two dimensional  open
string
gas at the same temperature $T'$. The number of species is now
simply $Q_1^2$, but the spatial size of the 2-branes in
the $y_2$ direction becomes ${\apm \over L_2'}$, which when combined with the
extra $T'$ gives us the same factor. 
The same consideration
applies to $y_3, y_4, y_5$ directions. So, there are
$Q_1^2 {\apm^4 T'^4\over L_2' L_3'
L_4' L_5'}$ species of open strings. Of course, from the 1-brane world
volume point of view, the masses of these open strings range from zero
to $T'$. But to the  accuracy of factors of order unity, we can treat
them as massless. So the entropy from 1-1 strings is  
\eqn\didvsii{
S_{11}  \sim \  Q_1^2 {\apm^4 T'^4\over L_2' L_3'
L_4' L_5'}  T' L_1' 
\sim {S_{bh} \over \ch^4\a_{min}}}
 
Now let us consider the 1-5  strings.
The 1-5 open strings have NN(Neumann-Neumann)  boundary
condition in the $y_1$ direction
and ND or DN boundary condition in the $y_2 ,
y_3, y_4, y_5$ directions. It is a
one dimensional gas. The ND or DN
boundary condition does not change the number
of species of the strings, so there are $Q_1 Q_5$ of them. 
 The entropy from the 1-5 strings
is then
\eqn\didvsiv{
S_{15} \ \sim \ Q_1Q_5  T' L_1'
\sim S_{bh} }

 The total entropy
$S_{open}$
and total local energy $E_{open}'$ 
carried the three kinds open strings are
\eqn\didvsetotal{\eqalign{
&S_{open} =  S_{11} + S_{55} + S_{15} \sim S_{15}   \sim S_{bh}  
 \cr
&E'_{open} \sim T' S_{open} \sim \Delta E_{bh}' 
}}
The relation between $E_{open}'$ and
$S_{open}$ just follows from the fact that we have an open string
gas at local temperature $T'$. The fact that $E_{open}'$ agrees with $\Delta
E_{bh}'$ justifies our initial assumption that the local temperature is
$T'$.  

We  see that at the matching point the entropy of the black hole
\didvmetric\ is carried by 1-5 strings, the contributions from 1-1 and
5-5 strings are suppressed by a factor of  ${1 \over \ch^4\a_{min}}$.
 This depends crucially on the deep throat geometry outside the
horizon. This result 
is consistent with that from precise counting \ghas\ and resolves a puzzle
in \ghjp. It was thought in \ghjp\ that at the matching point the metric
\didvmetric\ could be trusted all the way to the horizon and therefore a
much larger gravitational field dressing would exist. The local temperature
would therefore always be of order the string scale and it was puzzling
why the 1-1 and 5-5 strings did not contribute to the total entropy.
Indeed under this assumption the calculation shows that the entropy from
1-1 and 5-5 strings are of the same order as 1-5 strings (and they are
all of the same order of the black hole entropy). Now we see that the
gravitational dressing effect actually stops at $r \sim r^*$ so that
the local temperature is small in string units and the entropy contributions
from 1-1 and 5-5 strings are  suppressed.

\subsec{Six Dimensional Black Holes with D 0- and D 4-brane Charges}
The string metric of the six dimensional black hole with D 0- and D
4-brane charges is given by
\eqn\dodivmetric{\eqalign{
ds^2 = &(H_0 H_4 )^{1/2} \Bigl[ -(H_0 H_4 )^{-1} f dt^2
 + ( H_4)^{-1} (dy_1^2 +dy_2^2 +
 dy_3^2  +  dy_4^2) \cr 
&+ f^{-1}dr^2 +r^2 d\Omega_4^2 \Bigr] }}
where
\eqn\dodivhmfn{\eqalign{
H_i  &= 1+{r_0^3 \sinh^2\alpha_i \over r^3} \ \ (i=0, 4) \cr
f &= 1-{r_0^3 \over r^3} }}
The $y_a$ are coordinates
on the internal $T^4$. 
 The dilaton is
$e^{2 \phi} = H_0^{3/2} H_4^{-1/2}$. 
This black hole is a solution to the type
IIA string theory on $T^4$. 
 The  horizon is at $r=r_0$.
The extremal limit of the black hole can be obtained
by letting $r_0 \to 0$  while fixing $r_0^3 \sh^2\a_i$
 and has zero horizon area.
The extremal limit   preserves 1/4 of the
supersymmetries. 

The analysis and result of this black hole are similar to the previous
one.
The matching between the black hole  and 
 D-branes and strings occurs  when 
\eqn\dodivmpt{
r_0 \sim {  \apm^{1/2} \over (\ch\a_0 \ch\a_4)^{1/2}}}
                                                                    
Let  $\ch\a_{min}$ and $\ch\a_{max}$ be the smaller and
larger one of $\ch\a_0$ and $\ch\a_4$. 
The spatial geometry outside the horizon is the ``bottle neck": the
radius of the $S_4$ first
decreases, it reaches the  minimum at $r \sim r_0 (\ch\a_{min})^{2/3}$ and then increases.
The stringy behavior smears out the region
outside the horizon all the way  to the point where the curvature
becomes of order the string scale again.
Unlike the black hole considered in 3.1, the
gravitational field dressing will appear only when the hierarchy between
$\ch\a_0$ and $\ch\a_4$ is large enough, more precisely only when
 $\ch\a_{max} > \ch^3\a_{min}$.

Now let us consider the statistical description of the system. It should
be described by  $Q_0$ D 0-branes smeared in the 4-torus and $Q_4$ D
4-branes with some excitations. We will see that 
the   open  string gas picture gives consistent result.
 There are three
kinds of open strings: 0-0, 4-4 and 0-4  strings. We assume  they
live at 
 local temperature $T'$, which is   obtained from the Hawking
temperature and calculate the entropy  and energy carried by them. We
will see that they agree with  the black hole entropy and excess energy.
The calculation has to be done locally.

The  4-4 strings are a four dimensional gas. The 
entropy  is given by 
\eqn\dodivsiviv{
S_{44}  \sim \  Q_4^2
  T'^4 V' 
\sim \cases{{S_{bh} \over \ch^{12}\a_{4}}, &if $\ch\a_{0} >
\ch^3\a_{4}$;\cr {S_{bh} \over \ch^4\a_{0}}, &otherwise.\cr  }}
where $V'$ is the local volume of the 4-torus and
 $S_{bh}$ is the black hole entropy at the matching point. 
The 0-0 strings are related to the 4-4 strings by T-dualizing all the
internal directions in the 4-torus. To get $S_{00}$, we can simply
``T-dualize'' equation \dodivsiviv\
\eqn\dodivsoo{
S_{00}  
\sim \cases{{S_{bh} \over \ch^{12}\a_{0}}, &if $\ch\a_{4} >
\ch^3\a_{0}$;\cr {S_{bh} \over \ch^4\a_{4}}, &otherwise.\cr  }}
The 0-4 strings have 
 ND or DN boundary condition in the four internal 
 directions. 
 The ND or DN
boundary condition does not change the number
of species of the strings, so there are $Q_0 Q_4$ of them.
 The entropy from the 0-4 strings
is 
\eqn\didvsiv{
S_{04} \ \sim \ Q_0Q_4  
\sim S_{bh} }
The total entropy
$S_{open}$ and the total local energy $E_{open}'$ 
carried by the 3 kinds of  open strings is 
\eqn\dodivsetotal{\eqalign{
&S_{open} =  S_{00} + S_{44} + S_{04} \sim S_{04}   \sim S_{bh}
 \cr
&E'_{open} \sim T' S_{open} \sim \Delta E_{bh}'
}}
where $\Delta E_{bh}'$ is the local black hole excess energy.

So we  see that at the matching point, the entropy of the black hole
\dodivmetric\ is carried by 0-4 strings, the contributions from 0-0 and
4-4 strings are suppressed 
 by some powers in $\ch\a_{i}( i=0, 4)$.

\subsec{Four Dimensional Black Holes with D 2-brane,   D 6-brane and  NS
5-brane Charges}
The string metric of the four dimensional black hole with D 2-brane,   D
6-brane and NS  5-brane charges is given by
\eqn\diidvimetric{\eqalign{
ds^2 = &(H_2 H_6)^{1/2} H_5  \Bigl[ -(H_2 H_6 H_5 )^{-1} f dt^2
+ (H_2 H_6)^{-1}dy_1^2
 + ( H_2 H_6 H_5)^{-1} dy_2^2 \cr
+& (H_6 H_5)^{-1} (dy_3^2  +  dy_4^2 + dy_5^2 +dy_6^2) 
+ f^{-1}dr^2 +r^2 d\Omega_2^2 \Bigr] }}
where
\eqn\diidvidnsvhmfn{\eqalign{
H_i  &= 1+{r_0 \sinh^2\alpha_i \over r} \ \ (i=2, 5, 6) \cr
f &= 1-{r_0 \over r} }}
The $y_a$ are coordinates
on the internal $T^6$ which are identified with $L_a$.
 The dilaton is
$e^{2 \phi} = H_2^{1/2} H_6^{-3/2} H_5$. 
This black hole is a solution to the type
IIA string theory on $T^6$.
 The  horizon is at $r=r_0$.
The extremal limit of the black hole can be obtained
by letting $r_0 \to 0$  while fixing $r_0 \sh^2\a_i$
 and has zero horizon area.
The extremal limit   preserves 1/8 of the
supersymmetries. 
If we set the 5-brane charge $Q_5 = 0$, the black hole carries only the
two RR charges, for convenience this case will be included in the discussion of section 4. 
  In this section, we  consider the case when all  charges are large. 

The analysis and result for this black hole are similar to the previous
two black holes. The matching between the black hole  and 
  D-branes, NS 5-branes and strings occurs  at
\eqn\diidvimpt{
r_0 \sim {  \apm^{1/2} \over (\ch\a_2 \ch\a_6)^{1/2}\ch\a_5}}

Let  $\ch\a_{min}$  and $\ch\a_{max}$ be   the smallest and
the  largest   
 of $\ch\a_i (i=2, 5, 6)$.  
The spatial geometry outside the horizon has a deep throat from $r_0$ to
$r_0 \ch^2\a_{min}$. Note this will not happen if we set $Q_5 =0$. Adding NS 5-branes
balances the dilaton and gives us the deep throat. The whole throat is
smeared out by stringy behavior at the matching point. 
 If all $\ch\a_{i} $ are of the same order, 
 there is no large gravitational field
dressing;
if  $\ch\a_{max} >> \ch\a_{min}$, then there is still a
large dressing.

Now let us consider the statistical description of the system. It should
be described by  $Q_2$ D 2-branes,  $Q_6$ D
6-branes and $Q_5$ NS 5-branes with some excitations. The brane
configuration is as follows:
\eqn\diidviconfig{\eqalign{
D2 &: 1\ \ \ 2\ \ \  \ \cr
D6 &: 1\ \ \ 2\ \  \ 3\ \ \ 4\ \ \ 5\ \ \ 6 \cr
NS5 &: \ \ \ \ \ 2  \ \ \ 3\ \ \ 4\ \ \  5\ \ \ 6\ \     
}} 
We will see that
the   open string gas picture gives consistent result. There are three
kinds of open strings: 2-2, 6-6 and 2-6  strings. We assume  they
live at
 local temperature $T'$, which is   obtained from the Hawking
temperature and calculate the entropy  and energy carried by them. We
will see that they agree with  the black hole entropy and excess
energy. 

Before we do the calculation, we have to take care of  the  
  new feature brought in by the NS 5-branes. The 
5-branes are uniformly distributed in the $y_1$ direction. They  cut the
6-torus into $Q_5$ segments and 
each D 2- and D 6-brane  into
$Q_5$ pieces \jmas. In each  segment the $Q_2$ D-2 branes or
$Q_6$ D 6-branes can not be wrapped around the $y_1$ direction. We need
to know the local separation between adjacent 5-branes to see whether at
 temperature $T'$ the open strings can see this dimension or not.
Let $L_1$ and $L_1'$ be the asymptotic and
local size of the $y_1$ circle.
We have
\eqn\diidviqv{
Q_5 \sim {L_1r_0\sh2\a_5 \over \apm} \sim {L_1\ch\a_5 \over \apm^{1/2}
(\ch\a_2 \ch\a_6)^{1/2}} \sim {L_1' \over \apm^{1/2}}} 
where the second step follows from \diidvimpt\  and the relation between
$L_1$ and $L_1'$ is independent of whether there is large dressing or
not. 
 So the separation between adjacent 5-branes is always of order $\apm^{1/2}$.
 On the other hand, 
the  local temperature $T'$ is always much smaller than  the
 string scale. So, the $y_2$
direction is frozen for the open strings: the 2-6 and 2-2 strings become
an one dimensional gas, the 6-6 strings become a five dimensional gas. 
   
The entropy from the 6-6 strings is the sum of the
contributions
from each of the $Q_5$ segments
\eqn\diidvisvivi{
S_{66}  \sim \  Q_5 Q_6^2  T'^5 L_2' L_3' L_4' L_5' L_6'
\sim {S_{bh} \over \ch^4\a_{min}}}
where $L'_i (i=2, 3, 4, 5, 6)$ are the local sizes of the $y_i$ circles
and 
 $S_{bh}$ is the black hole entropy at the matching point.
The 2-2 strings are related to the 6-6 strings by T-dualizing the
$y_2$, $y_3$, $y_4$, $y_5$ directions. Since \diidvisvivi\  is  invariant
under this operation, we must have $S_{22}  \sim S_{66}$.
The entropy  from the 2-6 strings is 
\eqn\diidvisiivi{
S_{26}  \sim \  Q_5 Q_2 Q_6 L_2'T'
\sim {S_{bh} }}
The total entropy
$S_{open}$ and local energy $E_{open}'$
carried by the three kinds of  open strings are
\eqn\diidvisetotal{\eqalign{
&S_{open} =  S_{22} + S_{66} + S_{26} \sim S_{26}   \sim S_{bh}
 \cr
&E'_{open} \sim T' S_{open} \sim \Delta E_{bh}'
}}
where $\Delta E_{bh}'$ is the  local black hole excess energy.     

 So we  see that at the matching point, the
entropy of the black hole
\diidvimetric\ is carried by 2-6 strings, the contributions from 2-2 and
6-6 strings are suppressed by a factor of  ${1 \over
\ch^4\a_{min}}$.
 This
is consistent with the result from precise counting \hlm.

\newsec{Four Dimensional Black Hole with 4  RR Charges}
In this section we  consider a 
four dimensional black hole that carries 4 RR
 charges. We will go through the calculation in some detail because it
differs from black holes considered in section 3 in certain aspects.
 The string metric of the
black hole is 
\eqn\bhmetric{\eqalign{
ds^2 = &(H_1 H_2 H_3 H_4)^{1/2} \Bigl[ -(H_1 H_2 H_3 H_4)^{-1} f dt^2
+ (H_1 H_3)^{-1}  dy_1^2 + (H_1 H_4)^{-1} dy_2^2 \cr 
&+(H_1 H_2)^{-1} dy_3^2  + (H_2 H_4)^{-1} dy_4^2 + 
(H_2 H_3)^{-1} dy_5^2 + (H_3 H_4)^{-1} dy_6^2 \cr 
&+ f^{-1}dr^2 +r^2 d\Omega_2^2 \Bigr] }}
where 
\eqn\hmfn{\eqalign{
H_i  &= 1+{r_0 \sinh^2\alpha_i \over r} \ \ (i=1, 2, 3, 4) \cr
f &= 1-{r_0 \over r} }}
The $y_a(a=1, 2, 3, 4, 5, 6)$ are coordinates 
on the internal $T^6$ which are 
identified with length $L_a$.  The dilaton is
a constant.
This black hole is a solution to the type
IIB string theory on $T^6$. 
 The event horizon is at $r=r_0$. The extremal limit of the black hole
can be obtained by letting $r_0 \to 0$ while fixing $r_0 \sh^2\a_i$ and
has nonvanishing horizon area if all four charges are nonzero. 
The extremal limit 
 preserves 1/8 of the supersymmetries 
and can be 
thought of as  a superposition of four different kinds of 
extremal black 3-branes that are extended in the 
$(123), (345), (561), (246)$ directions separately and smeared out 
in the rest directions.  
  For the rest of this section, we will consider the case where at least
two charges are large.

The energy, excess energy, entropy, Hawking temperature and four RR charges of the black hole are
 \eqn\energy{\eqalign{
 E &\sim {L_1 L_2 L_3 L_4 L_5 L_6 \over g^2 \apm^4}\  r_0\
 (\ch 2\alpha_1 + \ch 2\alpha_2 +\ch 2\alpha_3 + \ch 2\alpha_4) \cr
 \Delta E_{bh} &\sim {L_1 L_2 L_3 L_4 L_5 L_6 \over g^2 \apm^4}\  r_0
  \( {1 \over \ch^2\alpha_1} + {1 \over \ch^2\alpha_2} +
  {1 \over \ch^2\alpha_3} + {1 \over \ch^2\alpha_4} \) \cr
 S_{bh} &\sim {L_1 L_2 L_3 L_4 L_5 L_6 \over g^2 \apm^4} \ r_0^2\  \ch 
\alpha_1 \ch\alpha_2 \ch\alpha_3 \ch\alpha_4 \cr
 T &\sim {1 \over  r_0 \ch\alpha_1 \ch\alpha_2 \ch\alpha_3 \ch\alpha_4} \cr
 Q_1 &\sim { L_4 L_5 L_6 \over g \apm^2} \ r_0 \sh 2\alpha_1 \cr
 Q_2 &\sim { L_1 L_2 L_6 \over g \apm^2} \ r_0 \sh 2\alpha_2 \cr
 Q_3 &\sim { L_2 L_3 L_4 \over g \apm^2} \ r_0 \sh 2\alpha_3 \cr 
Q_4 &\sim { L_1 L_3 L_5 \over g \apm^2}  \ r_0 \sh 2\alpha_4    }}
 where to  obtain 
$\Delta E_{bh}$  we have  used the relation that $\ch 2\a_i - \sh
2\a_i \sim {1 \over \ch^2\a_i}$ for any value of $\a_i$. 
 
The 
matching point is  
\eqn\traneqn{
r_0 \sim {  \apm^{1/2} 
\over (\ch\alpha_1 \ch\alpha_2 \ch\alpha_3 \ch\alpha_4)^{1/2}}}
At the matching point, the system should be 
described by the corresponding four kinds of D 3-branes 
with some excitations.  
The  D3-branes are wrapped around the internal $T^6$ 
in the following way:
\eqn\config{\eqalign{
&(D3)_1: 1\ \ \ 2\ \ \ 3\ \ \ \cr
&(D3)_2: \ \ \ \ \  \ \ \ \ 3\ \ \ 4\ \ \ 5\ \ \cr
&(D3)_3: 1\ \ \ \ \ \ \ \ \ \ \ \ \ \ \ \ \ 5\ \ \ 6\ \     \cr
&(D3)_4: \ \ \ \   2\ \ \ \ \ \ \ \ 4\ \ \ \ \ \ \ \ 6\ \ \cr
}}

There is a qualitative difference between the case where two or
three  charges are large and the case where all four  charges are large.
Let us discuss these two different cases in the next two subsections
separately.

\subsec{The Black Hole with Two or Three Large  Charges}

 Without losing generality, let us set $\a_4 = 0$, so $Q_4 = 0$ and the
black hole carries $Q_1, Q_2, Q_3$ only.
The 
D-branes have a large gravitational  field dressing
 and the metric
at the  the D-branes (i.e., $r \sim r_0$)  is 
\eqn\fixrmetric{\eqalign{
ds^2 \sim\  &( \ch\a_1 \ch\a_2 \ch\a_3)
 \Bigl[- {dt^2 \over  \ch^2\a_1 \ch^2\a_2 \ch^2\a_3}
+{dy_1^2 \over  \ch^2\a_1 \ch^2\a_3} \cr  
&+ {dy_2^2 \over
 \ch^2\a_1 } 
+ {dy_3^2 \over  \ch^2\a_1 \ch^2\a_2} +
{dy_4^2 \over \ch^2\a_2 } 
+{dy_5^2 \over  
 \ch^2\a_2 \ch^2\a_3} 
+
{dy_6^2 \over \ch^2\a_3 } \cr &+
{ \apm d\Omega_2^2 \over  \ch\a_1 \ch\a_2 \ch\a_3} \Bigr] }}
where we have substituted \traneqn\ into the metric. 
Let us use $T',\Delta E_{bh}'$ to 
denote the  local temperature and local energy, then
\eqn\rescale{\eqalign{
T' &\sim { 1 \over \apm^{1/2}} \cr
\Delta E_{bh}' &\sim {L_1 L_2 L_3 L_4 L_5 L_6 \over g^2 \apm^{7/2}}\cr  
}}

We will assume that the excitations are described by the massless open
strings and see  that this gives consistent result.  
There are six kinds of open strings in general:   
 three of them  end on the same kind of 
D-branes, the rest three stretch between 
different kinds of D-branes.
We    assume the local temperature to be 
${1 \over \apm^{1/2}}$ and 
 calculate the entropy and  energy carried by them.
 We will   see that  
they agree with the black hole entropy and excess energy.

 Let us use $i$-$j$ 
open strings to denote the open strings stretched
between the $i$th and $j$th kind $\bigl($which are labeled in
\config$\bigr)$ of D 3-branes 
and  $S_{ij}$ to denote 
the entropy contribution from them. There is an obvious symmetry 
among the three kinds of D 3-branes.  It is enough to 
just calculate $S_{11}$ and  $S_{12}$,  the rest can be obtained by the
symmetry.

We consider $S_{11}$ first. The 1-1 open strings form a three
dimensional gas.   
 The $Q_1$ D 3-branes are evenly distributed in the $y_4, y_5, y_6$
directions.
 So there are
$Q_1^2 {\apm^{3/2} \over L_4' L_5'
L_6'}$ species of open strings, for the same reason  explained in
section 3.1.  The entropy from the 1-1 strings is  
\eqn\SII{
S_{11} \ \sim \  Q_1^2 {\apm^{3/2} \over L_4' L_5'
L_6'} T'^3 L_1' L_2' L_3'
\sim S_{bh} \th^2\a_1}
where
\eqn\epmatch{
S_{bh} \sim {L_1 L_2 L_3 L_4 L_5 L_6 \over g^2 \apm^3}}
 is the black hole entropy at the matching point,
which follows from \energy\   and \traneqn.

Next we consider the 1-2 open strings. 
The 1-2 open strings have NN(Neumann-Neumann)  boundary 
condition in the $y_3$ direction, DD(Dirichlet-Dirichlet)  boundary
condition
 in the $y_6$ direction 
and ND or DN boundary condition in the $y_1 , 
y_2, y_4, y_5$ directions. It is a 
one dimensional gas. The ND or DN 
boundary condition does not change the number 
of species of open strings. The DD boundary
condition in the $y_6$ direction gives 
a factor of $\apm^{1/2} \over L_6'$ to
the number of species for the same
reason. The entropy from the 1-2 strings
is 
\eqn\SIII{
S_{12} \ \sim \ Q_1Q_2 {\apm^{1/2} \over L_6'} T' L_3' 
\sim S_{bh} \th\a_1 \th\a_2}
 
Because of the symmetry among the three kinds of 
D-branes, it is clear that
 the entropy from the $i$-$j$ open strings
is 
\eqn\Sij{
S_{ij} \sim S_{bh} \th\a_i \th\a_j}
The total entropy $S_{open}$ 
and local energy $E_{open}'$  carried by  these open strings are 
\eqn\Stot{\eqalign{
&S_{open} = \sum_{i, j=1}^3 S_{ij} \sim S_{bh} \sum_{i, j=1}^3 \th\a_i \th\a_j
\sim S_{bh} \cr
&E'_{open} \sim T' S_{open} \sim
\Delta E_{bh}'}
}
The sum $ \sum_{i, j=1}^3 \th\a_i \th\a_j$ in the entropy expression
includes six terms and takes value between 3 and 6 since we have assumed
at least two charges are large.

Now let us briefly discuss the two large charges case and three large charges case separately.
 If only   two  charges are large, we can ignore the third
charge.  We have two different kinds of D 3-branes that
intersect at a line, which is T-dual to the system of 0-branes and
4-branes     
 smeared out in  two more  dimensions or the system of 1-branes and
5-branes smeared out in  one more  dimension or the system of  2-branes
and
6-branes.
Up to T-duality, this black hole and the black holes considered in
section 3.1 and 3.2  are all the possible black holes that belong
to   category (B) that we classified in the introduction. The 
correspondence principle can be applied to all of them. 

If three  charges are large,  the entropy can also be
accounted for by massless open strings. However,   
  we have to remember that the precise counting
of entropy has already been done for 
a four dimensional near extremal black holes with three large charges
\hlm, the entropy was reproduced at weak coupling in the limit that the
excitations of the 4th kind of charge are much lighter than the other three
charges and form a dilute gas. 
 One might expect that similar precise counting can be done for this
black hole. This would suggest that the entropy is carried by  brane-antibrane excitations rather than open strings at
the matching point. We will do such a calculation in section 4.2. The
result is that  the entropy contribution from the 4th kind of branes and
antibranes is of the same order as that from the gas of the open
strings. But the gas of the brane-antibrane pairs is not dilute and the
calculation depends  on the assumption that the 4th kind of
branes and antibranes make up the entropy with the other three kinds of
branes independently.   
 If this assumption is correct, then a complete analysis would include both
types of excitations.  
 Which form the excitations really take should  depend  on   which type  
 gives the larger entropy.    

\subsec{The Black Hole with 4 Large RR Charges}
 One might wonder since the entropy of four dimensional extremal black hole
with four charges has been counted precisely \jmas \vbfl,
  why  we are still interested in this case. If the black hole is
 exactly at  the extremal limit, then 
 there is nothing new that needs to be said. If the black hole is only
near extremal with four large charges, we can still apply the
correspondence principle to it and ask   what the description 
  is at the
matching point.

Let  $r^{*} = r_0 \ch^2\a_{min}$, where $\ch\a_{min}$ is the smallest
of $\ch\a_i (i=1, 2, 3, 4 )$.  The spatial geometry outside the horizon
has a deep throat from $r_0$ to
$r^*$. We see the crucial difference  between this  case
and the previous case: turning on the 4th charge causes the
development of the deep throat, the length of the throat is 
determined by the smallest charge (by $\ch\a_{min}$, strictly
speaking).    
 The whole throat is smeared out by
stringy
behavior at the matching point.  If all $\ch\a_i$ are of the same order
there is no large
gravitational
field dressing;
if there is a hierarchy between the $\ch\a_i$'s,  then there is
still a
large dressing. In either case, we can write $H_i(r=r^*) \sim {\ch^2\a_i
\over
\ch^2\a_{min}}$. In the following, we will treat both cases
together. 

 The metric
at the D-branes (i.e., $r \sim r^*$) is
\eqn\branedress{\eqalign{
ds^2 \sim\  &\({\ch\a_1 \ch\a_2 \ch\a_3 \ch\a_4 \over \ch^4\a_{min}}\)
 \Biggr[- {dt^2 \over \({\ch\a_1 \ch\a_2 \ch\a_3 \ch\a_4 \over \ch^4\a_{min}
}\)^2} \cr
&+{dy_1^2 \over \({\ch\a_1 \ch\a_3 \over \ch^2\a_{min}}\)^2} 
+  {dy_2^2 \over
\({\ch\a_1 \ch\a_4 \over \ch^2\a_{min}}\)^2}
+ {dy_3^2 \over \({\ch\a_1 \ch\a_2 \over \ch^2\a_{min}}\)^2} +
{dy_4^2 \over \({\ch\a_2 \ch\a_4\over \ch^2\a_{min}}\)^2} \cr 
&+{dy_5^2 \over \({\ch\a_2 \ch\a_3 \over \ch^2\a_{min}}\)^2} 
+
{dy_6^2 \over \({\ch\a_3 \ch\a_4 \over \ch^2\a_{min}}\)^2} +
{ \apm d\Omega_2^2 \over \({\ch\a_1 \ch\a_2 \ch\a_3 \ch\a_4 \over \ch^4\a_{min}
}\)} \Biggr] }}
It  approaches the metric \fixrmetric\ smoothly as $\a_{min} \to 0$.

   Independent of whether there is large
dressing or not, the local temperature and local excess energy
are always given by   
\eqn\localte{\eqalign{
T' &\sim { 1 \over \apm^{1/2}\ch^2\a_{min}} \cr
\Delta E_{bh}' &\sim {L_1 L_2 L_3 L_4 L_5 L_6 \over g^2 \apm^{7/2} 
\ch^4\alpha_{min}}}}
  Compared with the black hole carrying less than four large charges,
 $T'$ and $\Delta E'_{bh}$ are suppressed by some power in ${1 \over
\ch\a_{min}}$. 

Another special feature about the 
four large charges case  is that most of the
entropy is the BPS entropy. We can imagine obtaining such a black hole
by adding excess energy to an extremal black hole with exactly the same
 charges (incidentally we notice
that if we add or take away some 
excess energy from  the system at the matching point, it
will
stay at the matching point, which does not happen for the less than four
large charges case). 
 The entropy of the extremal black hole is what we call the
BPS entropy and is given by 
\eqn\extremals{
S_{ex} = 2 \pi\sqrt{Q_1 Q_2 Q_3 Q_4}} 
The excess entropy $\Delta S_{bh}$ is defined to be the difference between
the total entropy and the BPS entropy
\eqn\deltaS{\eqalign{
\Delta S_{bh} \equiv S_{bh} - S_{ex} &= S_{bh}\(1- \sqrt{\tanh\a_1
\tanh\a_2\tanh\a_3
\tanh\a_4}\) \cr
 &\sim
{S_{bh} \over \ch^2\a_{min}}
}}
where we have used the property  
 $1 - \tanh\a_i \sim
{1\over \ch^2\a_i}$ for $\ch\a_i >> 1$.
At the matching point (or in the weak coupling limit for that
matter), the BPS entropy comes from the four kinds of D-branes \vbfl\ and
can be  counted  precisely. So what needs to be 
reproduced is the excess entropy $\Delta S_{bh}$. 

We will start by considering the open string gas picture and  see
that it fails to work. We then turn to the discussion of the
brane-antibrane picture (i.e.,
 the excitations take the form of brane and antibrane pairs), which is
a possible resolution. 
       
 We  will assume the local
temperature to be  $T'$ and calculate the 
entropy and energy carried by the massless open strings.
There are ten different kinds of open strings: four of them end on 
the same kind of D-branes, the rest six stretch between two different kinds
of D-branes. 

We consider $S_{11}$ first. 
 The 1-1 open strings should now have length of order
${\apm^{1/2} \over \ch^2\a_{min}}$.  The number of species of 1-1  
strings is 
$Q_1^2 {\apm^{3/2} \over L_4' L_5' L_6' \ch^6\a_{min}}$ and the 
entropy is 
\eqn\SIIst{
S_{11} \ \sim \  Q_1^2 {\apm^{3/2} \over L_4' L_5'
L_6' \ch^6\a_{min}} T'^3 L_1' L_2' L_3'
\sim {S_{bh}  \over \ch^{12}\a_{min}}}
Next, we consider  $S_{12}$. 
The number of species of 1-2 strings is now $Q_1Q_2 {\apm^{1/2} \over
L_6' \ch^{2}\a_{min}}$. The entropy from the 1-2 strings
is
\eqn\Sonetwo{
S_{12} \ \sim \ Q_1Q_2 {\apm^{1/2} \over L_6' \ch^2\a_{min}} T' L_3'
\sim {S_{bh}  \over \ch^4\a_{min}}  }
Because of the symmetry among the 4 kinds of D-branes, it is clear that
the entropy from the $i$-$j$ open strings is 
\eqn\SIJ{\eqalign{
&S_{ii} \sim {S_{bh}  \over \ch^{12}\a_{min}}\  (i=1, 2, 3, 4) \cr
&S_{ij} \sim {S_{bh}  \over \ch^{4}\a_{min}} \ (i,j=1, 2, 3, 4, i \ne j)
}}
The total entropy
and local energy  are
\eqn\Sopen{\eqalign{
&S_{open} = \sum_{i, j=1}^4 S_{ij} \sim {S_{bh} \over \ch^{4}\a_{min}} 
 \cr
&E'_{open} \sim T' S_{open} \sim
  {L_1 L_2 L_3 L_4 L_5 L_6 \over g^2 \apm^{7/2} \ch^{6}\a_{min}}}
} 
They are 
much smaller than the $\Delta S_{bh}$ and   
$\Delta E'_{bh}$. We can also do the alternative  calculation 
by putting all the excess energy $\Delta
E'_{bh}$ into the open strings. This  leads to  a  local
temperature of order ${1 \over \apm^{1/2}(\ch\a_4)^{4/3}}$, higher than
$T'$ 
 and an entropy of order  $S_{bh} \over (\ch\a_4)^{8/3}$, still much
smaller than $\Delta S_{bh}$.

 This shows that the open string gas picture
does not work. 
 If 
 the correspondence principle can be applied to this case in the usual context, we need
to search for other forms of excitations. 
In the following, we will consider the brane-antibrane
picture.
 
We consider the problem in the framework of canonical ensemble. The
D-brane system is immersed  in  temperature $T'$. Let us say creating a
certain number of brane-antibrane pairs costs local energy $ E_{pair}'$
and increases the entropy by $S_{pair}$. The increase in the energy
gives a Boltzman factor $e^{-{ E_{pair}' \over T'}}$, while the increase
in the entropy means the degeneracy of microstates is increased by
$e^{ S_{pair}}$ times. To find out how many pairs are excited, we need to
extremize $e^{-{ E_{pair}' \over T'} + S_{pair}}$ with respect to the
number of pairs. 

Since the system is near BPS, we can extremize with
respect to each kind of branes separately. Let us focus on the first kind
of D-brane and assume $Q_{\bar 1}$ pairs are excited.   
We  assume $ S_{pair}$ is given by 
\eqn\deltaS{
S_{pair} \sim  \sqrt{Q_{\bar
1}Q_2Q_3Q_4}},
 i.e., the $Q_1 + Q_{\bar 1}$ branes and $Q_{\bar 1}$
antibranes make up the entropy with the other three kinds of branes
independently.
The  $ E_{pair}'$ is given by
\eqn\deltaE{
 E_{pair}' \sim Q_{\bar 1} {L_1' L_2' L_3' \over g \apm^2}} 
Using \energy, \traneqn\ and \branedress\
        to extremize $e^{-{ E_{pair}' \over T'}+  S_{pair}}$ with
respect to  $Q_{\bar 1}$, we find 
\eqn\pairratio{
{Q_{\bar 1} \over Q_1} \sim {1 \over \ch^4\a_{min}}} 
Because of the symmetry among the 4 kinds of D-branes, we must also have 
\eqn\pairrationII{
{Q_{\bar 2} \over Q_2} \sim {Q_{\bar 3} \over Q_3} \sim {Q_{\bar 4}
\over Q_4} \sim {1 \over \ch^4\a_{min}}}  
It is easy to confirm that the entropy and excess energy
 contributions from each kind of D-branes
 are of the same order and the total entropy and excess energy  are
\eqn\deltase{\eqalign{
& S_{pair} \sim {S_{bh} \over \ch^2\a_{min}} \sim \Delta S_{bh} \cr
& E_{pair}' \sim \Delta E_{bh}'}}
We are using the same symbol $ S_{pair}$ and $ E_{pair}'$ here, hopefully
this will not cause any confusion.

This calculation depends critically on the assumption \deltaS.
 It was first shown in \hms\ that for black holes with multiple
charges,   one can formally define  brane and  antibrane numbers for each
charge
and write  the energy, entropy, etc.,  in terms of the  brane and
antibrane numbers. Take the black hole that we are considering  as a
example, the entropy can be written as
\eqn\paires{
S = 2 \pi \(\sqrt{Q_1}+\sqrt{Q_{\bar 1}}\)\(\sqrt{Q_2}+\sqrt{Q_{\bar
2}}\) \(\sqrt{Q_3}+\sqrt{Q_{\bar 3}}\)\(\sqrt{Q_4}+\sqrt{Q_{\bar
4}}\)
}
Moreover, if one fixes the charges and the excess energy of the black
hole,
 assumes the entropy is given by  \paires\ and maximizes
the entropy, one finds  the antibrane numbers to be the same as
 obtained from the black hole. What we have done is not exactly the
same in that it is not a asymptotic calculation but a local
calculation. It is interesting to see that it gives the right result.

We can do the same calculation for the three large charges case considered in
4.1. The number of pairs of the 4th kind of D-branes excited is 
\eqn\iiipair{
Q_{\bar 4} \sim { L_1L_3L_5 \over g \apm^{3/2} (\ch\a_1 \ch\a_2 \ch\a_3)^{1/2}}
}
 and the entropy from them is 
\eqn\iiipairs{
S_{pair} \sim \sqrt{Q_1
Q_2Q_3Q_{\bar4}} \sim S_{bh}}
We can estimate the  local density of these branes and
antibranes in the $y_1y_3y_5$ volume in the internal torus. 
It turns out to be of order ${1
\over g \apm^{3/2}}$, so it
is not a dilute gas.

\newsec{Eight Dimensional Black Hole with Two RR Charges}
In this section, we consider an eight dimensional near
extremal black hole that carries   two RR charges.
Unlike
the black holes considered in the previous sections,
the extremal limit of this black hole is what is
called non-marginal bound state. For a discussion 
of this kind of black holes, see \nojz. 
The string metric of the black hole is
\eqn\metricII{
ds^2 = H^{1/2} \Bigl[ -H^{-1} f dt^2+  \(1+(H-1)\cos^2\t\)^{-1} 
(dy_1^2 + dy_2^2) +
f^{-1}dr^2
+r^2 d\Omega_6^2 \Bigr]}
where
\eqn\fnIII{\eqalign{
H &= 1 + { r_0^5 \sh^2 \a \over r^5} \cr
f &=  1 - {r_0^5 \over r^5} \cr
  }}
 The dilaton is 
$e^{2 \phi} = H^{3/2} \(1+(H-1)\cos^2\t\)^{-1}$. The $\t$ is a constant
angle whose physical meaning will become clear. The $y_1, y_2$ are
coordinates on the internal $T^2$. 

This black hole is a solution to the
type IIA string theory
on $T^2$. The extremal limit  of this black hole was
obtained in \bmm.
The non-extremal solution can be constructed along
the same line. The  horizon is at $r= r_0$. In the metric
\metricII, the
extremal limit can be
obtained by letting $r_0 \to 0$ while fixing $r_0^5\sh^2\a$ 
 and it  preserves 1/2 of the
supersymmetries. If $\cos^2\t=0$, the solution corresponds to 
D 0-branes smeared out in the $y_1, y_2$ directions; if $\cos^2\t=1$, the
solution corresponds to
D 2-branes extended in the $y_1, y_2$ directions; if $0 < \cos^2\t
< 1$,  the solution interpolates between the two limiting cases
and carries both  0-brane charge ($\propto \sin\t$)     and  2-brane charge
($\propto \cos\t$). 

 To make the symmetry between the two charges more
explicitly and facilitate the calculation, it is better to consider the T-dual
 picture in type IIB.
 Let us 
T-dualize along the $y_2$ direction and still use $y_2$ as the 
coordinate,  the solution becomes
\eqn\metricIII{\eqalign{
ds^2 =\  &H^{1/2} \Bigl[ - H^{-1} f dt^2 + (H^{-1} \cos^2\t  +\sin^2\t)
d{y}_1^2 + (H^{-1} \sin^2\t  +\cos^2\t ) d{y}_2^2  \cr
\ & + 2 \sin\t \cos\t (H^{-1}-1) d{y}_1 d{y}_2 +
f^{-1}dr^2
+r^2 d\Omega_6^2 \Bigl] }}
The dilaton is $e^{2 \phi} = H$.
The ${y}_1, {y}_2$ are identified with length $L_1, L_2$. The metric 
looks a little bit complicated, however, if we perform a rotation in the
$y_1$-$y_2$ plane:
\eqn\rotation{
\pmatrix{y_1 \cr y_2}  = \pmatrix{\cos\t & -\sin\t \cr
\sin\t & \cos\t} \pmatrix{\tilde{y}_1 \cr \tilde{y}_2 }}
The metric becomes 
\eqn\metricii{
ds^2 = H^{1/2} \Bigl[ -H^{-1} f dt^2+  H^{-1} d\tilde{y}_1^2 + 
d\tilde{y}_2^2 +
f^{-1}dr^2
+r^2 d\Omega_6^2 \Bigr]}
which we recognize as the non-extremal black string solution in IIB
that carries
RR two form electric charge smeared out in the $\tilde{y}_2$ direction.
 Note that \metricIII\ and \metricii\
are physically different black holes: the rotation
does not make any difference locally in the ${y}_1$-${y}_2$
plane, but the global coordinate identifications
of
the 2-planes are different.

Now let us focus on solution \metricIII. The energy, excess energy, entropy, Hawking temperature and 
two RR charges of this black hole  are
\eqn\quan{\eqalign{
E &\sim {L_1 L_2 \over g^2 \apm^{4}}\  r_0^5 (\ch 2\a + {7 \over 5}) \cr
\Delta E_{bh} &\sim {L_1 L_2 \over g^2 \apm^{4}}\  r_0^5 \cr
S_{bh} &\sim {L_1 L_2 \over g^2 \apm^{4}}\  r_0^6 \ch\a  \cr
T &\sim { 1 \over r_0 \ch\a} \cr
Q_1 &\sim { L_2 \over g \apm^{3}}\  r_0^5 \sh 2\a \cos\t \cr
Q_2 &\sim { L_1 \over g \apm^{3}}\  r_0^5 \sh 2\a \sin\t \cr 
}}

The matching point is 
\eqn\matchpt{
r_0 \sim {\apm^{1/2} \over (\ch\a)^{1/2} }}
 The spatial geometry outside the horizon is the ``bottle neck" and is
the same as that of the black 2-brane with D 2-brane charge, so there is
no large gravitational field dressing. 
The system can be described by  a single
 D-string winding around the $T^2$ diagonally with an angle $\theta$ 
with the ${y}_1$ direction and massless open strings ending
on it.  The 
winding number along the ${y}_1$ and ${y}_2$ directions
 are $Q_1$ and
$Q_2$ respectively. The total  length  of the D-string  is
\eqn\coodlength{
L_{tot} = \sqrt{(Q_1 L_1)^2 + (Q_2 L_2)^2}  \sim  {L_1 L_2 \over g
\apm^{1/2} (\ch\a)^{1/2}} }
 Since $Q_1, Q_2$ are large numbers, we can think of the D-string
as wrapped around the $T^2$ uniformly. The  separation $d$ 
between adjacent turns is 
\eqn\sep{
d = {L_1 L_2 \over L_{tot}} \sim g
\apm^{1/2} (\ch\a)^{1/2} }

At temperature $T$,
an open 
string that starts from the  $i$-th turn of the D-string can
end on any other $j$-th turn   within $\apm T$ distance. Since 
there is only one D-string, the different species of open
stings are distinguished by $|i-j|$. So, the number of 
species  is ${\apm T\over d}$. The entropy $S_{open}$ and
the  energy $E_{open}$ 
of this gas of open strings are 
\eqn\entrp{\eqalign{
S_{open} &\sim {\apm T \over d} T L_{tot} \sim  {L_1 L_2 \over g^2
\apm \ch^2\a}  \sim S_{bh} \cr
E_{open} &\sim T S_{open} \sim {L_1 L_2 \over g^2 \apm^{3/2} (\ch\a)^{5/2} }
\sim \Delta E_{bh}}}
So, they agree with the black hole entropy and 
excess energy. 

In fact, without doing any calculation 
one can see that the correspondence principle should 
work for this black hole.
 As we have remarked, 
 this  black hole and the black hole described by \metricii\    are
 related by a rotation in the $T^2$. 
Locally in the $y_1$-$y_2$ plane, there is no difference between the two.
 The entropies per unit
area 
calculated in the open string gas
picture are the same and  the  areas of the two $T^2$ are the same, therefore 
the total entropies calculated in this way must be the same. 

If we go back to the original black hole \metricII, the picture
at the matching point
 is that the 0-branes are dissolved into the 2-branes and
become magnetic flux on the 2-branes,   they form a bound state \gl.
 
These black holes can obviously be generalized. One can start 
with a near extremal black p-brane in type II
 that carries a certain kind of RR charge,
smear this black p-brane out in several other dimensions, 
make arbitrary rotations in  the internal torus  and finally 
T-dualize some of the internal directions if one wants to. All black
holes with two RR charges in the  category (A) that we classified in the
introduction are included in this kind of constructions. 
The entropy of such black holes can all be understood by the 
correspondence principle.

\newsec{Discussions}

In this paper we have considered  the entropy  of  near extremal
black holes with multiple charges in the context of the correspondence principle \ghjp\ of
Horowitz and Polchinski. 
 We have  considered all black holes with two RR charges
in the category (A) and (B) that are classified in the introduction and black
holes with three or four
 RR charges. We have also considered a black hole with two RR
charges and NS 5-brane charge. We found that  the
black hole entropy can always be counted for by the massless open
strings on the D-branes for all cases except one. The exception is 
a four dimensional black hole
with four RR charges, in which   case the open string gas picture does not work
because it contributes too small an entropy. We turned to the idea of
brane-antibrane excitations for this case and found they contribute
 the correct entropy. But the
calculation depends on 
the assumption that the branes and antibranes make up the entropy with
the other three kinds of branes independently. The precise counting of
entropy has been done for some of the black holes considered in this
paper. Our results from the correspondence principle are consistent with 
the precise counting for those cases.

 We have seen that the correspondence principle can be used to relate
states of a large class of black holes to states of strings and
D-branes. This  can be thought of as a statistical understanding  
of the Bekenstein-Hawking entropy for black holes that have size of
the string scale.  
It does not mean that we   
 have a statistical understanding of the  
entropy for  black holes 
that are much larger than the string scale. 
We do not yet understand why the black hole entropy does not
change when we change the string coupling constant while  keep
the  
horizon area fixed in Planck units, i.e., why the black hole
entropy is proportional to the horizon area measured in Planck units.

\vskip 1cm
 \centerline{\bf Acknowledgments}

We would like  to thank G. Horowitz for helpful discussions and
reading of manuscript. We would also like to thank R. Emparan,
 J. Pierre and J. Polchinski for helpful discussions. This work was
supported in part by NSF Grant PHY95-07065. 

\listrefs

\end